\def\year{2021}\relax
\documentclass[letterpaper]{article} %
\usepackage{aaai21}  %
\usepackage{times}  %
\usepackage{helvet} %
\usepackage{courier}  %
\usepackage[hyphens]{url}  %
\usepackage{graphicx} %
\usepackage{natbib}  %
\usepackage{caption} %
\usepackage{xcolor}
\usepackage{booktabs}
\usepackage{amsmath}
\usepackage{multirow} %

\usepackage{graphicx}
\usepackage{subcaption}

\title{\datasetname: a Multi-modal Dataset of Election Fraud Claims on Twitter}
\author {
    Anton Abilov\textsuperscript{\rm 1,2},
    Yiqing Hua\textsuperscript{\rm 1,2},
    Hana Matatov\textsuperscript{\rm 3},
    Ofra Amir\textsuperscript{\rm 3},
    Mor Naaman\textsuperscript{\rm 1,2}\\
}
\affiliations {
    \textsuperscript{\rm 1} Cornell Tech \\
    \textsuperscript{\rm 2} Cornell University \\
    \textsuperscript{\rm 3} Technion \\
    aa2776@cornell.edu, 
    yiqing@cs.cornell.edu,
    hanama888@gmail.com,
    oamir@technion.ac.il,
    mor@jacobs.cornell.edu
}

\newcommand{\smidge}{{\kern .05em}}

\def\imagebox#1#2{\vtop to #1{\null\hbox{#2}\vfill}}

\newcommand\datasetname{\emph{VoterFraud2020}}
\newcommand\detractor{\textit{detractor}}
\newcommand\promoter{\textit{promoter}}

\newcommand{\bnm}{\begin{newmath}}
\newcommand{\enm}{\end{newmath}}
\newcommand{\bne}{\begin{newequation}}
\newcommand{\ene}{\end{newequation}}

\newenvironment{newmath}{\begin{displaymath}%
\setlength{\abovedisplayskip}{4pt}%
\setlength{\belowdisplayskip}{4pt}%
\setlength{\abovedisplayshortskip}{6pt}%
\setlength{\belowdisplayshortskip}{6pt} }{\end{displaymath}}

\newenvironment{newequation}{\begin{equation}%
\setlength{\abovedisplayskip}{4pt}%
\setlength{\belowdisplayskip}{4pt}%
\setlength{\abovedisplayshortskip}{6pt}%
\setlength{\belowdisplayshortskip}{6pt} }{\end{equation}}

\newcommand{\secref}[1]{Section~\ref{#1}}

\newcommand{\figref}[1]{Figure~\ref{#1}}
\newcommand{\tabref}[1]{Table~\ref{#1}}

\newcommand{\verylongrightarrow}[1]             %
      {\setlength{\unitlength}{.01in}           %
      \begin{picture}(#1,1) \put(0,0){\vector(1,0){#1}} \end{picture}}

\newlength{\saveparindent}
\newlength{\saveparskip}
\newcounter{ctr}

\begin{document}

\maketitle

\begin{abstract}
The wide spread of unfounded election fraud claims surrounding the U.S. 2020 election had resulted in undermining of trust in the election, culminating in violence inside the U.S. capitol.
Under these circumstances, it is critical to understand the discussions surrounding these claims on Twitter,
a major platform where the claims were disseminated. 
To this end, we collected and released the \datasetname~dataset, a multi-modal dataset with~7.6M tweets and~25.6M retweets from~2.6M users related to voter fraud claims.
To make this data immediately useful for a diverse set of research projects, we further enhance the data with cluster labels computed from the retweet graph,
each user's suspension status, and the perceptual hashes of tweeted images. 
The dataset also includes aggregate data for all external links and YouTube videos that appear in the tweets. 
Preliminary analyses of the data show that Twitter's user suspension actions mostly affected a specific community of voter fraud claim promoters, and exposes the most common URLs, images and YouTube videos shared in the data. 
\end{abstract}

\section{Introduction}

Free and fair elections are the foundation of every democracy.
The 2020 presidential election in the United States was probably one of the most consequential and contentious such events. 
Two-thirds of the voting-eligible population voted,
resulting in the highest turnout in the past 120 years~\cite{record}.
The Democratic Party candidate Joe Biden was elected as the president.
 
Unfortunately, efforts to delegitimize the election process and its results were carried out before, throughout and after the election \cite{CenterforanInformedPublic2021}. 
Claims of voter fraud, most of them unfounded~\cite{superspreader}, were spread through public statements by politicians,
by the media, and on social media platforms.
As a result, 34\% of Americans say that they do not trust the election results as of December, 2020~\cite{trustpoll}.
While prior research had shown that allegations of widespread voter fraud in the recent 20 years is not supported by credible evidence~\cite{goel2020one}, claims of voter fraud have shown to have a significant negative impact on confidence in electoral integrity~\cite{Berlinski}.  
Indeed, on January 6th, 2021,
believing that the election was `stolen',
mobs breached the U.S. capitol while the Congress voted to certify Biden as the winner of the election. %
In theory and in practice, unsubstantiated voter fraud allegations have great ramifications for the integrity of the election and the stability of democracies in the U.S. and beyond. 

Social media platforms like Facebook, Twitter, YouTube and Reddit play a significant role in political events~\cite{vitak2011s,allcott2017social},
and the 2020 election was no exception~\cite{ferrara2020characterizing}.  
In particular, Twitter has been the focus of public and media attention as a prominent public square where ideas are adopted and claims---false or true---are spread~\cite{vosoughi2018spread,grinberg2019fake}.
It is thus important to understand the participants, discussions, narratives, and allegations around voter fraud claims on this specific platform.

In this work,
we release \datasetname, a multi-modal Twitter dataset of 7.6M tweets and 25.6M retweets that are related to voter fraud claims.
Using a manually curated set of keywords (e.g., ``voter fraud'' and ``\#stopthesteal'') that was further expanded using a data-driven approach, 
we streamed Twitter activities between October 23rd and December 16th, 2020.
We performed various validations on the limits of our stream, given Twitter's API constraints~\cite{Morstatter2013StreamingCoverage}, and estimate that 
we were able to retrieve around~60\% of the data containing our crawled keywords.

We further enhanced the \datasetname~dataset
in order to make it accessible for a broader set of researchers and future research: 
(1) We cluster users into communities according the network of their retweet patterns and release the community labels for each user; (2) Given Twitter's widespread post-election suspension action, we crawl and include each users' suspension status as of January 10th, 2021; 
(3) We compute and share the perceptual hashes of 168K images that appeared in the data;
(4) We aggregate and share metadata about 138K external links that appeared in the tweets, including 12K unique YouTube videos.
Our dataset also allows researchers to calculate the amount of Twitter interactions with the collected tweets, users, and media items,
including number of retweets and quotes from various clusters, or from suspended users.

A preliminary analysis finds a significant cluster of users who were promoting the election fraud related claims,
with nearly 7.8\% of them suspended in January.
The suspensions focused on a specific community \textit{within} the cluster.
A simple analysis of the distribution of images, based on visual similarity, exposes that the most broadly shared (by number of tweets) and the most retweeted images are different.
Although recent research has shown that voter fraud claims are pushed mainly by mass media~\cite{benkler2020mail},
we also find that external links referenced by promoters of the claims point mostly to low-quality news websites, streaming services, and YouTube videos.
Some of the most widespread videos claiming `evidence' of voter fraud were published by surprisingly small channels. 
Most strikingly, all of the top ten channels and videos spreading voter fraud claims were still available on YouTube as of January 11th, 2021.

We believe that the release of \datasetname, the largest public dataset of Twitter discussions around the voter fraud claims, with the enhanced labels and data, will help the broad research community better understand this important topic at a critical time.

\section{Data Collection}

Our data collection process involved streaming Twitter data using a data-driven manually curated set of keywords and hashtags. We report on the span and volume of the collected data, as well as on analyses estimating its coverage.

\subsection{Streaming Twitter Data}

We used a data-driven approach to generate a list of keywords and hashtags related to election fraud claims in an iterative manner.
We started with a single phrase and two derived keywords: \texttt{voter fraud} and \texttt{\#voterfraud}. We first used a convenience sample of 11M political tweets consisting of the tweets of 2,262 U.S. political candidates and the replies to those tweets, collected between July 21st and Oct 22nd, 2020 using the Twitter Streaming API~\cite{twitterapi}.
We then identified hashtags that co-occur with  our meta-seed keywords, \texttt{voter fraud} and \texttt{voterfraud}.
We selected all hashtags that appeared in at least 10 tweets and co-occurred with either of the meta-seed keywords at least 50\% of the time. 
From the resulting set, we manually filtered out those that were not directly relevant to voter fraud.
To this end, two members of the research team reviewed the hashtags, including, if needed, searching for them on Twitter to see whether they produce relevant results.
Only the hashtags that were agreed on by both evaluators were added, resulting in an initial set of hashtags that was added to the two original keywords.

To find more related hashtags, we computed the Jaccard coefficient between each of our seed hashtags and all other hashtags that appeared in the new stream.
We added to our set all hashtags that had a Jaccard coefficient greater than 0.001 with any of the seed hashtags.
Three members of the team again reviewed this list by 1) excluding  hashtags that were not related to voter fraud,
2) adding keywords corresponding to the hashtags~(e.g. \texttt{\#discardedballots} corresponds to \texttt{discarded ballots}), and
3) adding relevant hashtags or keywords that the researchers observed while searching for hashtags from the generated list.
Both the seed list and the final list of keywords and hashtags we used for streaming are included in the Appendix (\tabref{tab:hashtags}).

We collected data using the Twitter streaming API~\cite{twitterapi}.
The \datasetname~dataset includes tweets from 17:00, October 23rd, 2020 to 13:00 December 16th, 2020.
We expanded the keywords list on Oct. 31st with additional keywords,
and added \texttt{\#stopthesteal} as it started trending on November 3rd.
While streaming, we stored each tweet's metadata (e.g., user ID, text, timestamp). We also downloaded all image media items included in the tweets.
In total, we collected 3,781,524 original tweets, 25,566,698 retweets, and 3,821,579 \textit{quote tweets} (i.e. tweets that include a reference to another tweet). 
Note that quote tweets are included in the Twitter stream when either the new tweet or the referenced (quoted) tweet include one of the keywords or hashtags on the list.
In total, we collected tweets from 2,559,018 users who posted, shared or quoted one or more tweets with these keywords.

\subsection{Coverage Analysis}
\label{sec:coverage}
Since the Twitter streaming API provides only a sample of the tweets, especially for large-volume keywords~\cite{Morstatter2013StreamingCoverage}, we performed multiple tests to evaluate and estimate the coverage of the \datasetname~dataset. This analysis suggests that the dataset covers over~60\% of the content shared on Twitter using the keywords we tracked.

\paragraph{Retweet and Quote Coverage.}
We evaluated the coverage of retweets and quote tweets by comparing the counts of these objects in the stream to Twitter's metadata.
When a new retweet for an original tweet appears in the stream,
the API returns the tweet's metadata including the current retweet count and quote count of the original tweet.
In other words, if an original tweet $t_i$ is retweeted, it will appear in the stream as a retweet $rt_j$, and the metadata for $rt_j$ will include the total number of retweets of $t_i$ so far.
From this metadata, it is easy to define the \textit{retweet coverage} as the ratio of the total number of retweets ($rt$ objects) streamed and stored in our dataset,
over the sum of all retweet counts of the original $t$ tweets, returned by the API in the latest $rt$ retweet of each original tweet.
The quote coverage is defined analogously.
According to this analysis, the \datasetname~dataset captured 63.2\% of the retweets and 62.6\% of the quote tweets.
These findings compare favorably with previous work that shows a single API client captures only 37.6\% of the retweets through the Streaming API~\cite{Morstatter2013StreamingCoverage}.

\paragraph{Comparison with \textit{\#Election2020}.}
To further evaluate the coverage on the voter fraud tweets, we compared our dataset with a previously published Twitter dataset of the U.S. 2020 election~\cite{chen2020election2020}.
The creators of the \textit{\#Election2020} dataset used the streaming API to track 168 keywords that are broadly related to the election and 57 accounts that are tied to candidates running for president. 

As in \datasetname, the keyword `voter fraud' was also used to collect data for \textit{\#Election2020}. We used this overlap to estimate our coverage. 
First, we can directly compare the relative volume and overlap between the `voter fraud' tweets in both datasets.
We expect our \datasetname~to have a higher volume of such tweets because of its more focused set of keywords. Second, if we assume sampling for both streams is independent and random, we could estimate the coverage of \datasetname~by looking at the proportion of \textit{\#Election2020} tweets that also appear in our data.

To this end, we extracted all tweets and retweets that contain this keyword from both datasets
posted on two days following the November 3rd election data: November 6th and November 13th.
The analysis, performed on December 17th, was limited to two days as we had to obtain the content of the tweets of the \textit{\#Election2020} dataset by ``hydrating'' them (i.e. using the tweet IDs to get the full tweet text using the Twitter API).
We were unable to hydrate the full data, presumably due to inactive accounts and deleted tweets.
The hydration yielded 92.4\% of the \#Election2020 data from November 6th (a total of 1.4M tweets/3.5M retweets), and 91.1\% of the data from November 13th (1.3M tweets/3M retweets).

\begin{figure}[t!]
    \centering
    \includegraphics[width=8cm]{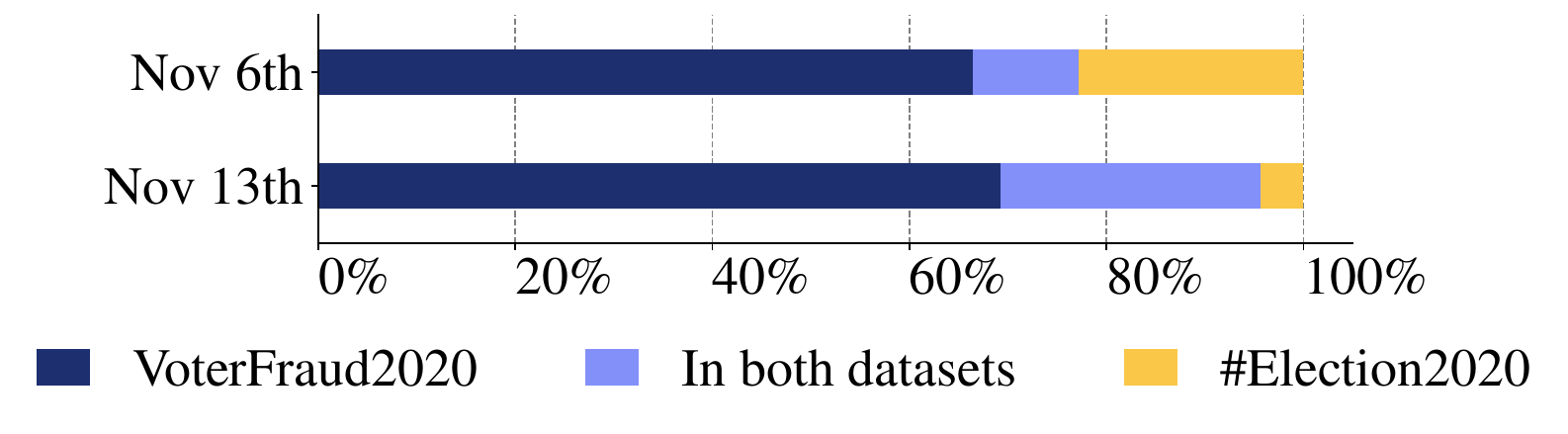}
    \caption{Coverage comparison between our dataset and \textit{\#Election2020} for tweets containing `voter fraud'. }
    \label{fig:coverage_comparison-fig}
\end{figure}

In total, our \datasetname~data includes 45,322 `voter fraud' related tweets on November 6th, 2.3 times as much as recorded in \#Election2020.
The ratio is even higher on November 13th, when we obtained 47,313 tweets, 3.1  times as much as in \textit{\#Election2020}.
\figref{fig:coverage_comparison-fig} breaks down the coverage by dates~(separated by rows),
in the two datasets~(by different colors).
From left to right,
the bars show the percentages of tweets that are available only in our dataset~(dark blue),
that are available in both datasets~(light blue),
and that are available only in \textit{\#Election2020}~(yellow).
On any given day, 
the \datasetname~dataset contains substantially more tweets related to voter fraud,
especially when the estimated total volume is lower.
On November 13th~(second row),
\datasetname~ contained 95.7\% of the combined data~(left two bars) while \textit{\#Election2020} only collected 30.7\%~(right two bars) of the tweets.
These numbers also indicate that \datasetname's sample includes 32.1\% of the related samples in \textit{\#Election2020} on November 6th and~85.9\% on November 13th.
We acknowledge that these two numbers are not consistent, presumably because of November 6th's much higher volume of activity. If these samples are indeed independent, though, it means that our lower bound of coverage is November 6th's 32.1\%.

Based on these coverage analyses,
we conclude that \datasetname~is, at the time of submission, the largest known public Twitter dataset of voter fraud claims and discussions.

\section{Data Enhancement}

To ensure the reusability of our data,
we took the following steps to enhance the raw streaming data.
We performed a community analysis of users according to the retweet graph, and release the community labels.
Given Twitter's large-scale suspension of accounts and the public interest in those actions, we also queried Twitter for each user's status on 10th of January,
and share the user status as active/not-found/suspended.
Furthermore,
to enable research on spread of images and visual misinformation,
we encode all images shared in the tweets with perceptual hash that allows for easy comparison and retrieval of similar content in the data.
Finally,
we release the set of URLs that appeared in the dataset,
as well as the YouTube metadata for each YouTube video URLs.

\paragraph{Retweet Graph Communities.}
Retweet networks have been frequently analyzed in previous work in order to understand political conversations on Twitter~\cite{arif2018acting, cherepnalkoski2016retweetnetwork}. 
Using community detection algorithms,
researchers are able to study key players, sharing patterns, and content on different sides of a discussion surrounding a heated political topic.  

To compute these communities, we first constructed a retweet graph of the \datasetname~dataset, where nodes represent users and directed edges correspond to one user (the target of the edge) retweeting another (the source).
Edges are weighted according to the number of times the corresponding source user has been retweeted by the target user. 
The resulting network consists of 1,887,736 nodes and 16,718,884 edges. 

To detect communities within the graph,
we used the Infomap community detection algorithm~\cite{bohlin2014mapequation},
which captures the flow of information in directed networks. 
Using the default parameters, the algorithm produces thousands of communities. 
By excluding all communities that contain fewer than 1\% of the nodes we are left with 90\% of all nodes\footnote{Since the graph only includes retweeting and retweeted users, this number corresponds to 73.8\% of all users in our dataset.} which are clustered into five communities (see \tabref{tab:community-detection}). 

\begin{figure}
\centering
\tiny
\begin{subfigure}[t!]{0.33\textwidth}
    \centering
    \includegraphics[width=\textwidth]{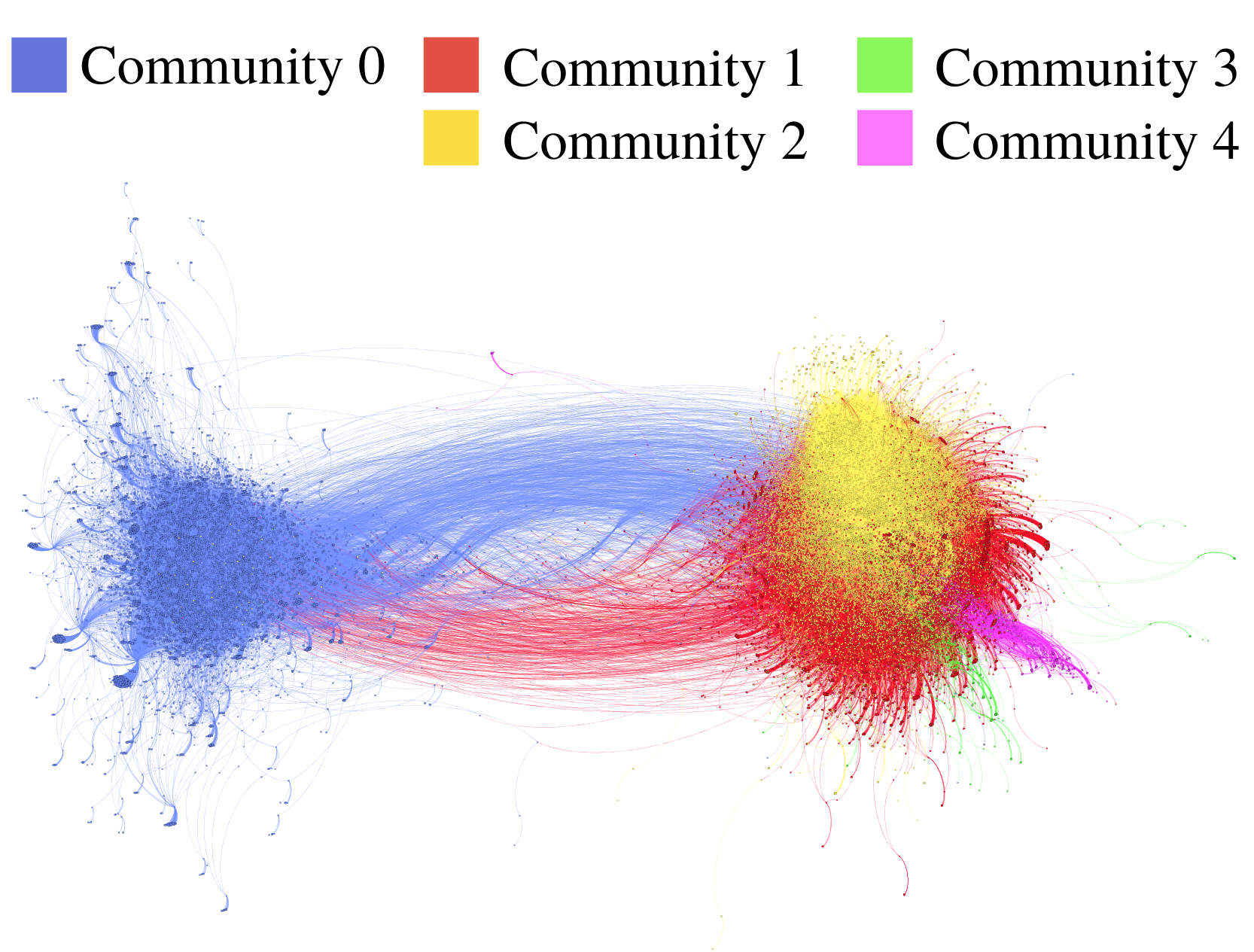}
    \caption{\label{fig:infomap_clustering}}
\end{subfigure}
\qquad
\begin{subfigure}[t!]{0.33\textwidth}
    \centering
    \includegraphics[width=\textwidth]{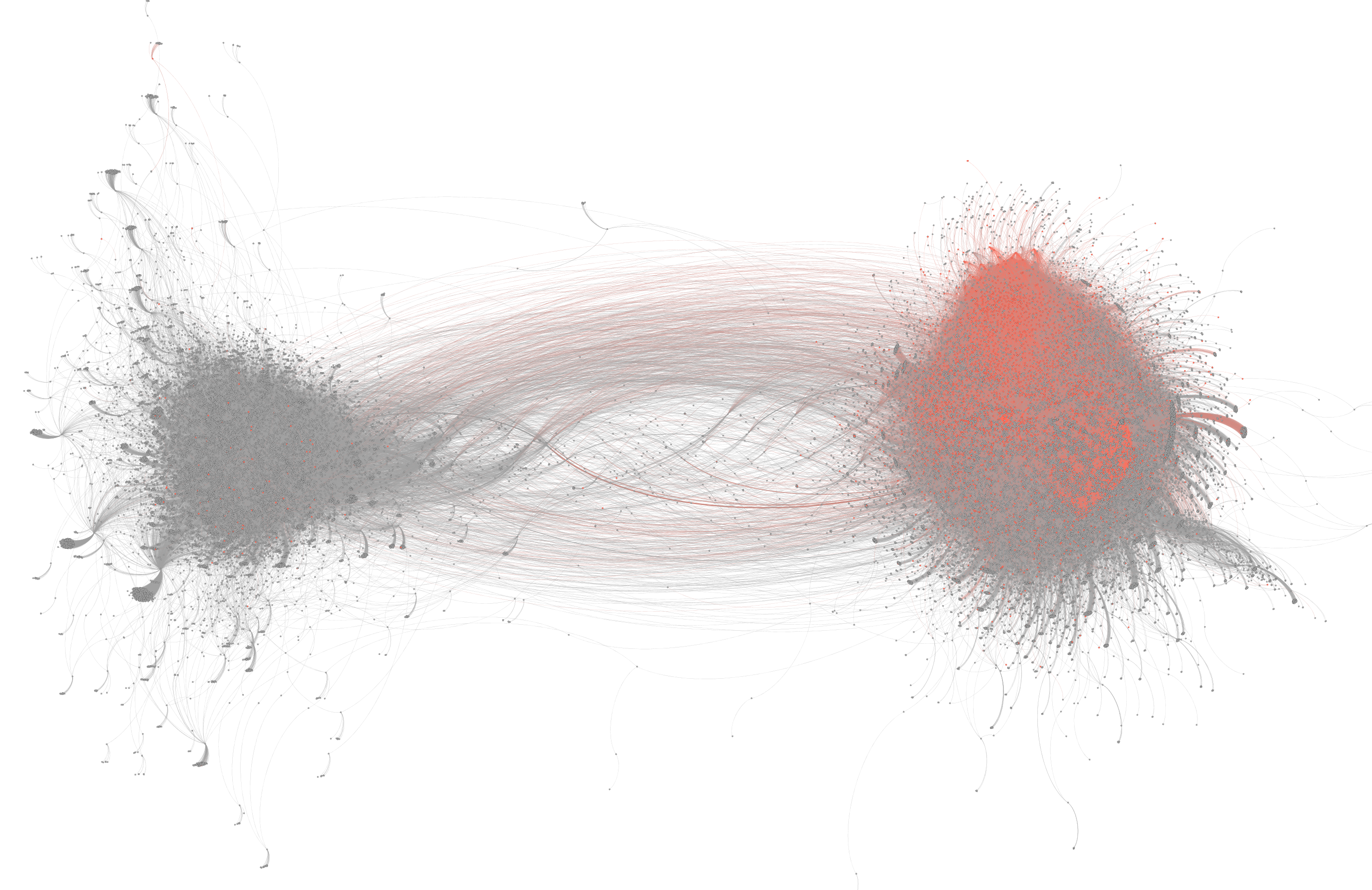}
    \caption{\label{fig:infomap_clustering_suspended}}
    
\end{subfigure}
    \qquad
\caption{\label{fig:retweet_graph}
{\tabular[t]{@{}l@{}}(a) Retweet graph colored by communities. \\ 
(b) Suspension status~(orange: suspended users).
\endtabular}
}
\end{figure}

In \figref{fig:infomap_clustering},
we visualize the retweet network using the Force Atlas 2 layout in Gephi~\cite{bastian2009gephi},
using a random sample of 44,474 nodes and 456,372 edges.
The nodes are colored according to their computed community as described in \tabref{tab:community-detection}. Edges are colored by their source node.
The visualization indicates that the nodes are split between two distinct clusters: community 0~(blue) on the left and communities~1, 2,~3 and~4 on the right. 
By examining the top users in each community,
we conclude that community~0 mostly consists of accounts that tend to refute and \textit{detract} from the voter fraud claims,
while the communities on the right consist of accounts that \textit{promote} the voter fraud claims.
For brevity,
in the following analyses,
we refer to the cluster on the left as the \detractor~cluster,
and the cluster with community 1,2,3,4 on the right as the \promoter~cluster.

\begin{table}[t!]
\centering
    \fontsize{9}{10}\selectfont
    \begin{tabular}{lrrrrrr}
        \toprule
        Community & Users
        & \% of users & Tweets & Images \\ 
        &&suspended && tweeted \\ 
        \midrule
	0 & 860,976 & 1\% & 1,199,587 & 30,506 \\
	1 & 437,783 &  4.6\% & 644,219 & 14,423 \\
	2 & 342,184 & 14.1\% & 3,982,990 & 94,115 \\
	3 & 33,857 & 1.5\% & 27,699 & 867 \\
	4 & 23,414 & 1.6\% & 24,191 & 753 \\
        \bottomrule
    \end{tabular}
    \caption{\label{tab:community-detection} Community statistics.}
\end{table}%

Community~2 is more deeply embedded in the \promoter~cluster compared to Community~1,
as we observe tweets from Community~1 being retweeted by Community~0 on the left, but not from Community~2.
Most of the tweets from in our data are written in English (not surprisingly, as we were tracking English language keywords),
except for users in Community~3 who mainly post  tweets in Japanese and users in Community~4 who write mostly in Spanish.
We include the list of top five Twitter accounts in each community by the number of community retweets in the Appendix.

Note that due to the partisan nature of the U.S. politics,
most of the \promoter~users are likely aligned with right-leaning politics, and \detractor~users align with left-leaning politics.
By inspecting the top 10 retweeted domains in each cluster (\tabref{tab:topdomains}), and correlating the sources with political alignment evaluations by AllSides\footnote{allsides.com} and Media Bias/Fact Check\footnote{mediabiasfactcheck.com}, we confirm that 90\% of the top news sources shared by the \promoter~cluster are right-leaning, and 90\% of the top news sources shared by the \detractor~cluster are left-leaning.

To identify users that are prominent within each of these two clusters, we calculate the closeness centrality of the user nodes in each cluster.
In a retweet network this metric can be interpreted as a user's ability to spread information to  other users in the network \cite{okamoto2008closeness}.
We compute the top-k closeness centrality to find the 10,000 most central nodes within the \detractor~and \promoter~clusters \cite{bisenius2017closeness}.

The dataset includes users' community labels and their closeness centrality score in their cluster (\detractor~and \promoter~clusters).
We also include additional community-based metrics for the tweet: \textit{retweet count by community $X$} and \textit{quote count by community $X$}.
For a tweet $t_i$,
the \textit{retweet count by community $X$} is the total number of retweets $rt_i$ it received from each user $u_X$ in community $X$.
The metric is also computed, analogously, for quote tweets.

\paragraph{Labeling Suspended and Deleted Users}
\label{sec:enhancement_inactive}

When the electoral college was set to confirm the election results on January 6th, 2021, the allegations of voter fraud took a dramatic turn, which culminated in the storming of the US Capitol. 
Subsequently, Twitter took a harder stance on moderating content on their platform and suspended at least 70,000 accounts that were engaged in propagating conspiracy theories and sharing QAnon-content~\cite{twitter2021safety}.
This ban has substantial implications for researchers seeking to understand the spread of voter fraud allegations on Twitter, since the Twitter API does not allow the ``hydration'' of Tweets from suspended users. 
In order to understand the distribution of suspensions in our dataset we queried the updated user status of all users in our dataset on January 10th, a few days following the ban. 
The Twitter API returns a user status that indicates if the user is active, suspended or not found (presumably deleted).
In total, 3.9\% of the accounts (99,884 accounts) in our data were suspended.

In \figref{fig:infomap_clustering_suspended}, we color the nodes in the randomly sampled retweet graph according to each user's suspension status (suspended users in orange). The visualization shows how Twitter's suspension efforts primarily targeted users within the \promoter~ cluster. In our data, we find that 88.3\% of the suspended users that were part of the top five communities were part of this cluster. Moreover, the figure shows that the suspensions greatly overlap with Community~2 
in \figref{fig:infomap_clustering}. The Data Analysis section below provides further details about this overlap and the suspended users.

We enhance the \datasetname~dataset by labeling tweets and users that were suspended. 
This metadata will enable research on the suspensions, and ease hydration of the data from Twitter by allowing hydraters to skip content that is no longer available. 
Relatedly, we also include two additional metrics for each tweet: \textit{retweet count by suspended users} and \textit{quote count by suspended users}. 

Due to its immense public interest, we have retained the \textit{full} data we retrieved from the 99,884 suspended users including 1,240,405 tweets and 6,246,245 retweets. This detailed data is not part of \datasetname. However, we will distribute an anonymized version of this data to published academic researchers upon request.

\paragraph{Images.}
\label{sec:enhancement_image}
Because of their persuasive power and ease of spread, there is a growing interest in analyzing how visual misinformation spreads both within a platform or across platforms~\cite{zannettou2018origins,highfield2016instagrammatics,paris2019deepfakes, moreira2018image, zannettou2020characterizing}.
However, visual information such as images or videos is difficult for many researchers to study due to computational and storage costs.
Here, we make the information about image content shared in \datasetname~easier to use by researchers by sharing \textit{perceptual hash values} for these images \cite{petrov2017wavelet,zauner2011rihamark}. 
Common perceptual hashes are binary strings  designed such that the Hamming distance~\cite{zauner2011rihamark} between two hashes is close if and only if the two corresponding images are \textit{perceptually} similar.
In other words, an image that is only slightly transformed,
for example, by re-sizing, cropping, or rotation,
will have a similar hash value to the original image.

With these numeric hash values, researchers can easily find duplicates and near-duplicate images in tweets, without working directly with cumbersome image content.
To this end, we download all image media items that were posted in the tweets in the streaming data,
and encode them with three different types of perceptual hashes. 
As the definition of perceptual similarity is often subjective and the underlying algorithms are often different,
various hash functions have different performance characteristics dealing with various types of image transformations.
Therefore, we encode the images in our dataset with \textit{three} perceptual hash functions: the Perceptive Hash~(pHash), the Average Hash~(aHash), and the Wavelet Hash~(wHash)~\cite{petrov2017wavelet,zauner2011rihamark}.

In total, our streamed tweets included 201,259 image URLs,
167,696 of them were successfully retrieved during streaming.
We provide some more details about the distribution of these images in Section~\ref{sec:analysis}.

\paragraph{External Links.}
\label{sec:enhancement_links}
Misinformation campaigns are known to use broad cross-platform information, often via links to other sites~\cite{wilson2020cross,golovchenko2020cross}.
Hence, we extracted and publish the set of external (non-Twitter) URLs that were referenced in the tweets.
For ease of use, we resolved URLs that point to a redirected location~(e.g. \url{bit.ly} URLs) to their final destination URL. %
Our streamed tweets included references to a total of 138,411 unique URLs,
appearing in 609,513 tweets. 

Since a large portion (over 12\%) of all URLs in the data were YouTube links, we further enhanced the data with YouTube-specific metadata.
A key motivation for this specific focus was the known role YouTube plays generally in spreading misinformation~\cite{hussein2020measuring,papadamou2020just} and specifically its role in the 2020 election and voter fraud claims~\cite{youtubefraud,youtubefraud1}.
For each YouTube video that was shared in the collected tweets, 
we used YouTube's Data API~\cite{youtubeapi},
to retrieve the video's title, description, as well as the id and title of the channel that posted it. We retrieved the YouTube metadata on Jan 1st, 2021. On that data, out of the 13,611 unique video ids that we have queried, 1,608 were no longer available resulting in 12,003 YouTube URLs with full additional metadata.

\section{Data Sharing and Format}
\label{sec:sharing}

Our \datasetname~dataset is available for download under FAIR principles~\cite{wilkinson2016fair} in CSV format\footnote{\url{https://figshare.com/account/projects/96518/articles/13571084}}.
The data includes ``item data'' tables for tweets, retweets, and users keyed by Twitter assigned IDs and augmented with additional metadata as described below.
The data also includes the list of images that appear in the dataset, 
indexed by randomly generated unique IDs.
Finally, the data includes aggregated tables for URLs and for YouTube videos including the information described in~\secref{sec:enhancement_links}.
The dataset tables and the fields they contain are summarized on Github\footnote{\url{https://github.com/sTechLab/VoterFraud2020}}.

The \datasetname~dataset conforms with FAIR principles.
The dataset is \emph{Findable} as it is publicly available on Figshare, with a digital object identifier (DOI): 10.6084/m9.figshare.13571084.
It is also \emph{Accessible} since it can be accessed by anyone in the world through the link.
The dataset is in csv format, hence it is \emph{Interoperable}.
We release the full dataset with descriptions detailed in this paper,
as well as an online tool to explore the dataset at \url{https://voterfraud2020.io},
making the dataset \emph{Reusable} to the research community.

The tables for Tweets and Retweets
contain the full set of items that were collected, 
including from suspended users.
These tables do not include raw tweet data beyond the ID, according to Twitter's ToS.
However, to support use of the data without being required to download (``hydrate'') the full set of tweets, we augment the Tweets table with several key properties. 
For each tweet we provide the number of total retweets as computed by Twitter~(\texttt{retweet/quote\_count\_metadata}),
as well as the number of retweets and quotes we streamed for this tweet from users in each of the five main communities~(\texttt{retweet/quote\_count\_community\_X}, X ranging from 0 to 4).
Note that the latter do not add up to the Twitter metadata retweet count due to the coverage issues listed in \secref{sec:coverage}. 
The Tweet table properties also include the \texttt{user\_community} (0--4) for the user who posted the tweet,  
computed using methods listed in \secref{sec:enhancement_inactive}.
Some of the Twitter accounts were not clustered into one of the five main communities.
In this case,
the \texttt{user\_community} label is \texttt{null}.
With this augmentation, researchers using this dataset could very quickly, for example, select and then hydrate a subset of the most retweeted tweets from non-suspended users in Community~2. 
As the tweet itself and the ID of the user who tweeted it is not available until hydration,  Twitter's users' privacy is preserved.

The Users table is similarly augmented with aggregate information about the importance of the user in the dataset, including the community that they belong to, their centrality in the two meta-clusters,
    \detractor~and \promoter~
    (\texttt{\small{closeness\_centrality\_detractor\_cluster}} and \texttt{\small{closeness\_centrality\_promoter\_cluster}}),
and the amount of attention (retweets and quotes) they received from other users in the different communities.
We also note whether, according to the data we collected, the user had been suspended.
With this data, interested researchers can quickly focus their attention and research on the main actors in each community. 

The Images table includes a list
of all the image media items retrieved in the stream, their unique media ID, and the ID of the tweet in which the image was shared. 
We augment this table with the image hash using three types of perceptual hash functions -- 
aHash, pHash and wHash, as detailed in \secref{sec:enhancement_image}.
This augmentation, together with the link to the Tweet ID, will allow researchers to quickly identify and hydrate popular images using the tweet metadata.
Researchers also have the option of identifying matches in the data for images that come from any other source, by computing and comparing the perceptual hash values for new query images. 

The two aggregate tables---the URLs table and the YouTube Videos table--- provide information about the popularity of each of these objects in the dataset: aggregate retweet and quote counts for the URL or video link, both using the Twitter metadata and the count of objects in our stream from the various communities.
In addition, these tables are augmented with metadata about the item (URL or YouTube video) as noted in \secref{sec:enhancement_links}.

\begin{figure}
    \centering
    \includegraphics[width=8cm]{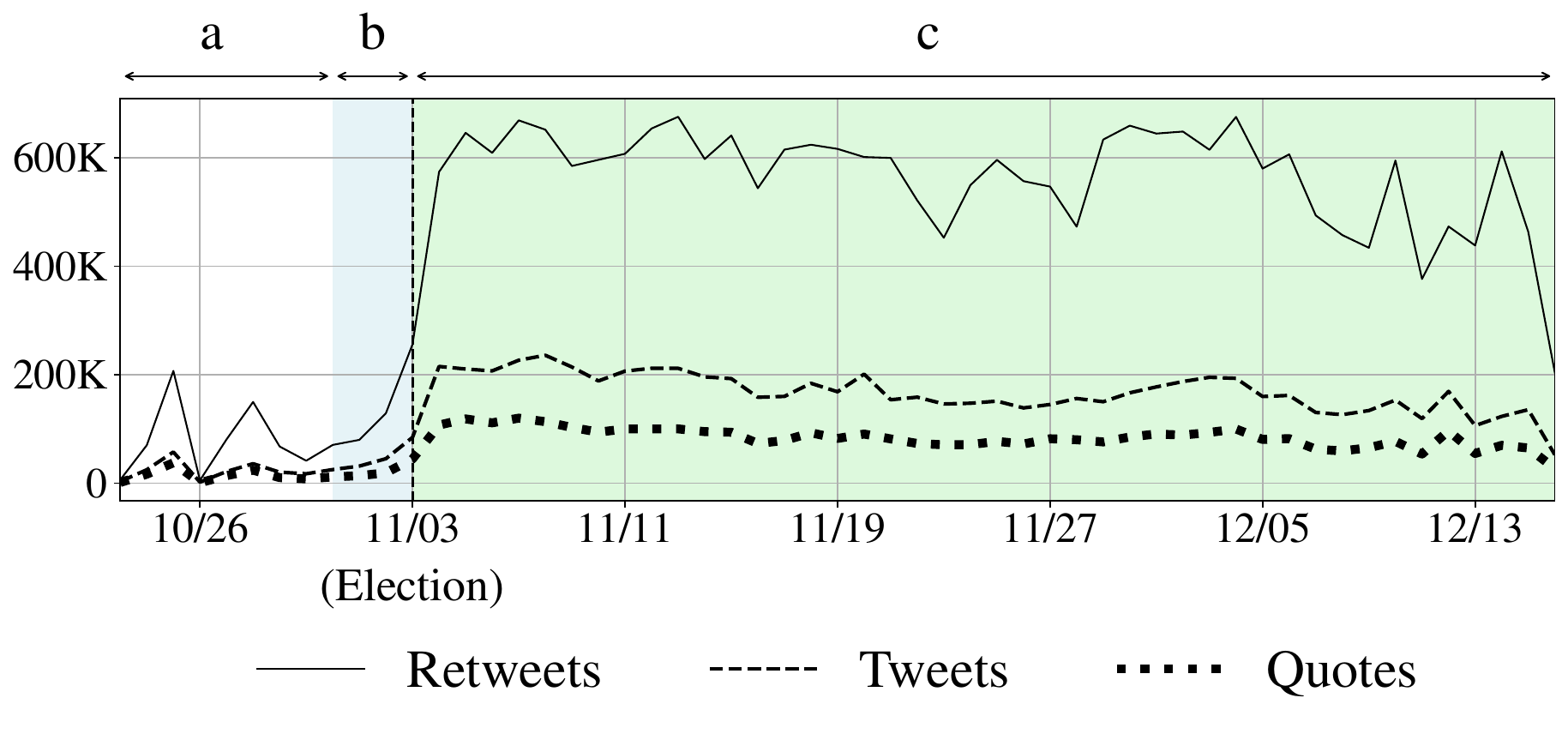}
    \caption{Temporal overview of the dataset showing number of streamed tweets, quotes and retweets per day. The shaded regions mark the expansions of the keyword set. 
    } %
    \label{fig:tweets_over_time}
\end{figure}

\begin{figure*}[t!]
\centering
\fontsize{9}{10}\selectfont
\begin{tabular}{cccccc} 
\multicolumn{2}{c}{(a)} &
\multicolumn{2}{c}{(b)} &
\multicolumn{2}{c}{(c)}\\
\multicolumn{2}{c}{\includegraphics[height = 1.2in]{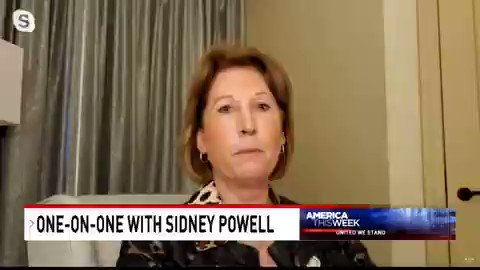}} &
\multicolumn{2}{c}{\includegraphics[height = 1.2in]{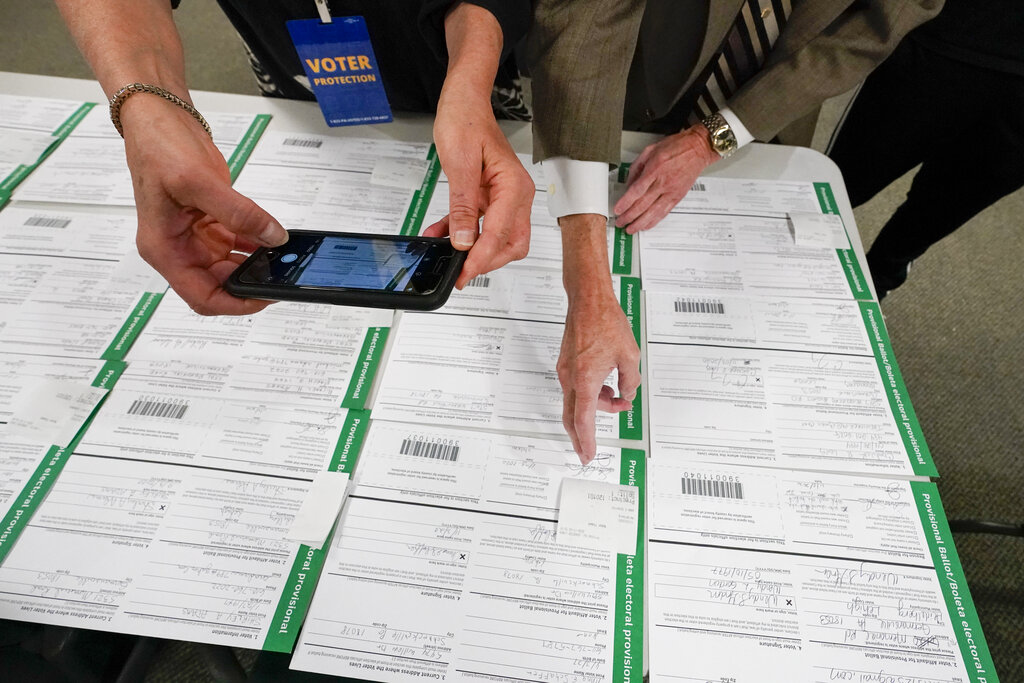}} &
\multicolumn{2}{c}{\includegraphics[height = 1.2in]{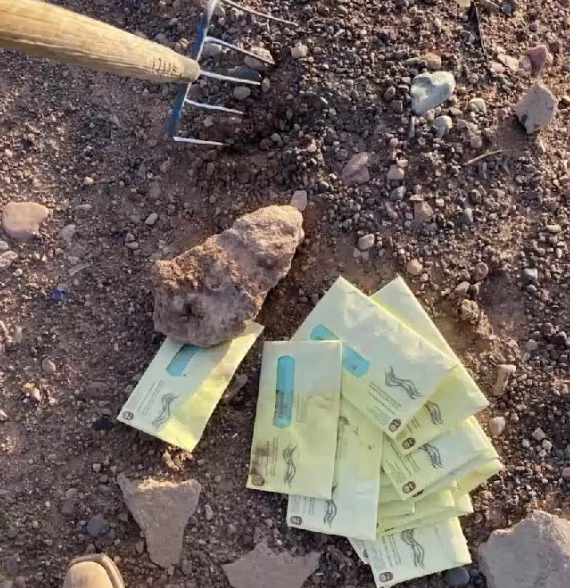}}\\
Tweets & Retweets~(by promoters) &
Tweets & Retweets~(by promoters) &
Tweets & Retweets~(by promoters) \\
15 & 18,020 &
11 & 10,424 &
34 & 10,250 \\

\end{tabular}
\caption{
Top three most retweeted images in the~\promoter~cluster: (a)--(c),
with the number of unique tweets they appeared in and the number of retweets by users in the \promoter~cluster.
Image (c) was cropped to fit the figure.}
\label{fig:visual_example_clusters}
\end{figure*}

\section{Data Analysis}
\label{sec:analysis}
We performed a preliminary analysis of our dataset and its different modalities -- tweets and users, images, external links -- to demonstrate its potential interest and provide some initial guiding insights about the data.

\paragraph{Tweets and Users.}
\figref{fig:tweets_over_time} shows the amount of retweets~(solid), original tweets~(dashed) and quote tweets~(dotted) in the \datasetname~dataset over the time~(X-axis) of the data collection.
Three shaded regions, from left to right,
mark the expansion of our set of keywords on October 31st~(light blue, region b) and November 3rd~(light green, region c). 
The Y-axis specifies the daily count. 
In general, except for the large increase after the election date~(November 3rd, dotted vertical line),
the volume of the stream remains roughly the same.
On average, there are 170,938 tweets, 576,136 retweets, and 85,488 quote tweets per day after the election.

Our manual inspection shows that top tweets retweeted by the \detractor~cluster often condemn the alleged voter fraud claims,
while top tweets on the \promoter~cluster indeed make voter fraud claims.
Not surprisingly, among the top ten most retweeted tweets in the \promoter~cluster,
nine were tweeted by President Trump.
We refer readers to our project website for more details about popular tweets.

While the \promoter~cluster seems rather homogeneous (\figref{fig:infomap_clustering}), users in Community~2 (yellow) stand out in both their level of activity and the rate in which they were suspended. 
Community~2 was highly active in our dataset. For example, 
Community 2 comprises 18.1\% of the users,
but contributed 68\% of the \datasetname~tweets,
and 74\% of the retweets.
Moreover, 14\% of Community~2's users were suspended by Twitter by the time we collected the account status data as described above, a much higher rate than the other communities, as shown in \tabref{tab:community-detection}. 
In total, Community~2 was responsible for 46.1\%
of all suspensions amongst the users we associated with the top five communities.
The suspension effect, and its focus on Community~2, can also be observed in \figref{fig:infomap_clustering_suspended}, when compared visually against \figref{fig:infomap_clustering}.

Promoters of the QAnon conspiracy theory were heavily involved in spreading unsubstantiated voter fraud claims~\cite{Tollefson2021}. We conduct preliminary analysis to evaluate the QAnon presence in the \datasetname~dataset and in the suspensions, based on early reports that suggested Twitter’s suspensions were targeting QAnon users \cite{Conger2021}. We curated a set of QAnon-related hashtags\footnote{The set of QAnon hashtags used in our analysis: \#awake, \#cabal, \#calmbeforethestorm, \#cbts, \#enjoytheshow, \#greatawakening, \#neonrevolt, \#outoftheshadows, \#patriqts, \#pizzagate, \#q, \#qanon, \#qmovie, \#qproofs, \#savethechildren, \#stqrm, \#thegreatawakening, \#theshow, \#thestorm, \#wga, \#wwg, \#wwg1wga}, and identified 50,385 users associated with QAnon by counting the users that have either tweeted or retweeted one of the hashtags, or mentioned the hashtag in their profile description. Of these users we find that 52.4\% users have been suspended as of January 10th, providing indication that Twitter’s suspensions focused on the QAnon community. 
We find that for these QAnon users for whom we had network data, 82.7\% were part of community 2, where suspension rates were highest.
Further, the rate of QAnon hashtags in Community 2 was 5 to 99 times higher than other communities in the retweet graph.

A full analysis of the suspended accounts and their network communities, and the potential impact of the suspension is out of scope for this dataset paper, but can be easily performed using the data we share in \datasetname. 
For example, the data shows that 
35\% of the \promoter~cluster users that were retweeted more than 1,000 times (1,596 in total) were suspended. %

To conclude, our preliminary analysis shows that alleged election fraud claims mostly circulate in the \promoter~cluster, and in particular in the most active Community~2 within the cluster.
The recent moderation efforts from Twitter seem to have affected this active community,
and did \textit{not} broadly target all accounts involved in promoting such claims.
There is some indication that Community~2 has significant overlap with followers of the QAnon conspiracy theory.

\subsubsection{Images.}
We conducted a preliminary examination of matching and repeated images in \datasetname~to analyze the distribution of images related to voter fraud claims.
Our data, using the perceptual hash functions described in Section~\ref{sec:enhancement_image}, allows tracking of duplicate and near-duplicate images that were posted in  multiple tweets. 
In this analysis, we experimented with three perceptual hash functions and refer to two images as matching if they have an \textit{identical} perceptual hash value. 

\begin{figure}[t!]
\centering
\begin{subfigure}[b]{0.46\textwidth}
    \centering
    \imagebox{4.5cm}{
    \includegraphics[width=\textwidth]{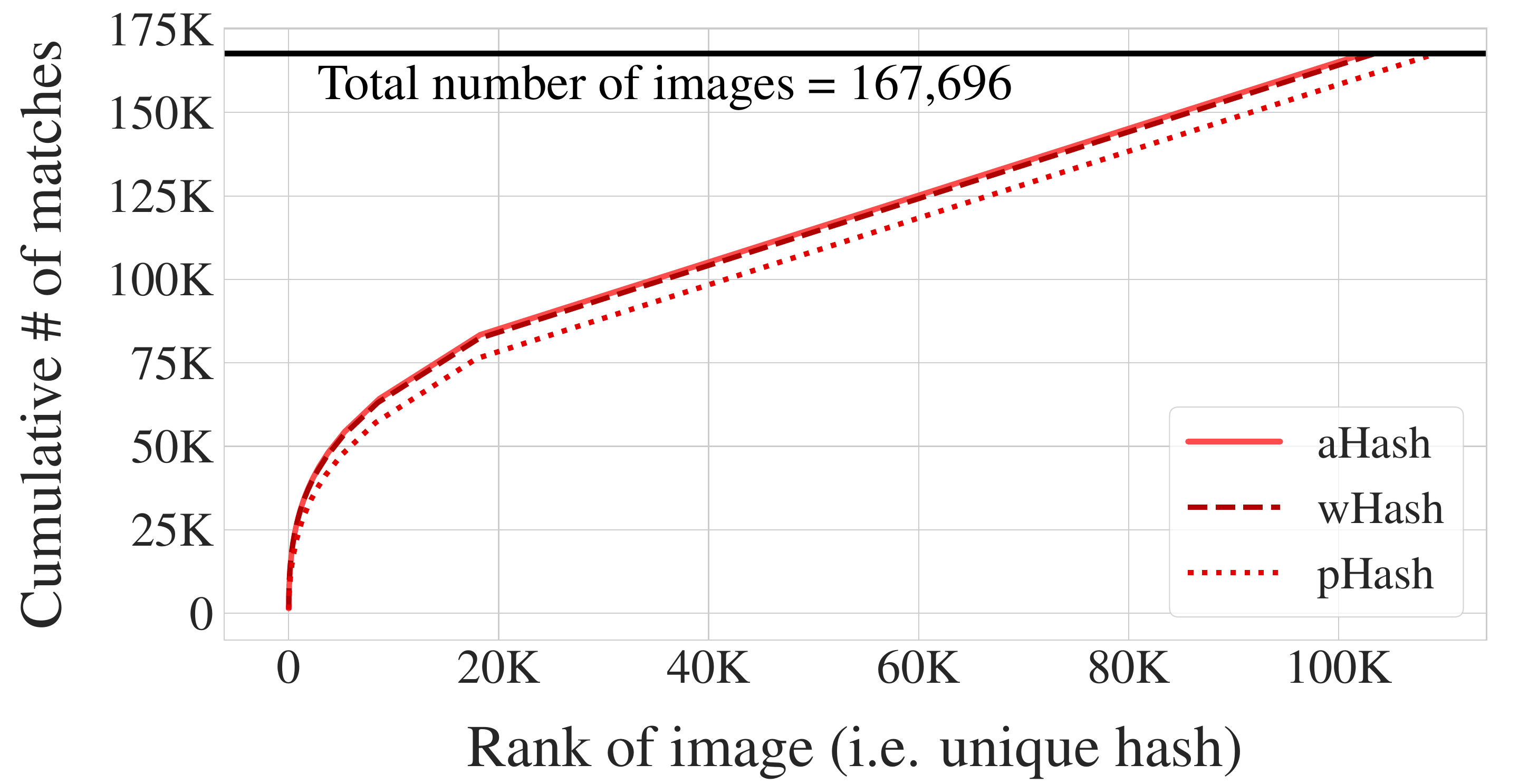}}
    \caption{\label{fig:image_cumulative_plot}}
\end{subfigure}
\begin{subfigure}[b]{0.25\textwidth}
    \centering
    \imagebox{3.5cm}{
    \includegraphics[width=\textwidth]{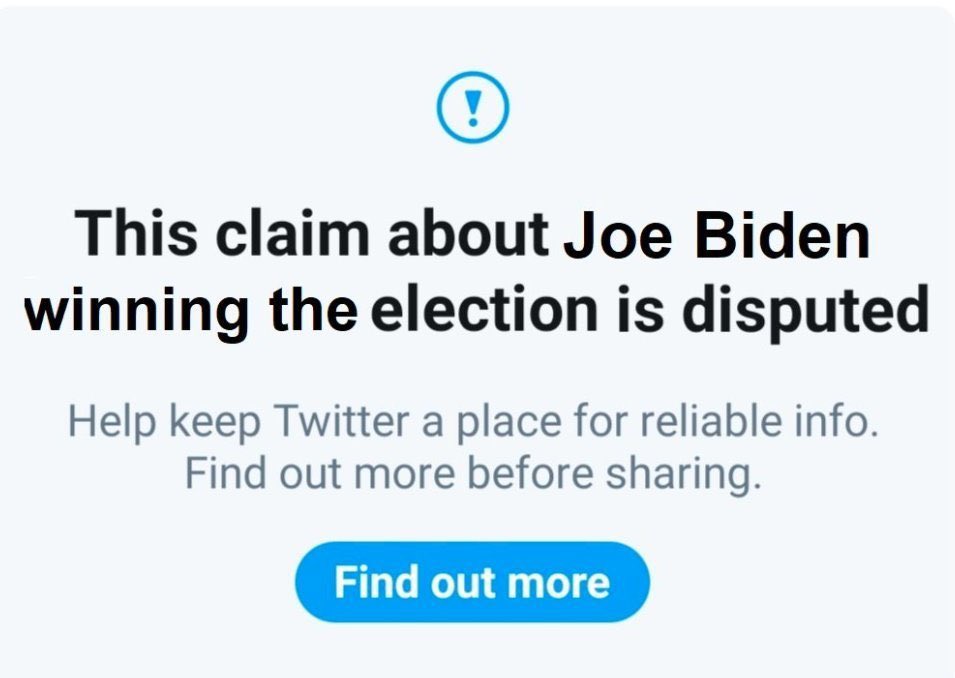}}
    \caption{\label{fig:max_appearances_image}}
\end{subfigure}
\caption{\label{fig:images_analysis}
(a) The cumulative number of repeated images by hash matches.
(b) The most tweeted image.}

\end{figure}

For example, 
there are 109,312~(out of 167,696) images with the same pHash value. 
Of these, 17,831 were shared in more than one tweet, an average of 4.27 times. 
In other words, 34\% of the image instances in~\datasetname~tweets appear in more than one tweet. 
Figure~\ref{fig:max_appearances_image} presents the image that appeared in the most unique tweets: the image (based on the perceptual hash value) appeared in over 1,000 tweets, according to all three hash functions. 

Figure~\ref{fig:image_cumulative_plot}
shows the cumulative distribution of the number of unique perceptual hashes in~\datasetname~(Y-axis),
with hash values sorted based on the number of unique tweets in which they appear, from the highest to the lowest (X-axis). 
For example, according to pHash, the 1,000 images shared in the largest number of unique tweets appeared together in 25,019 different tweets (not including retweets).
Although in general the results are similar when using different hash functions,
pHash is the most ``conservative'' of the three in terms of assigning matches.
Overall, 
our results are similar when using different hash functions.

We further investigate the popularity of images defined by the number of retweets
within the~\promoter~cluster.
After grouping images by the same pHash value,
we present in Figure~\ref{fig:visual_example_clusters} the top three images that have been retweeted in the~\promoter~cluster.
Also note that despite the high values of cluster retweets,
all these ``popular'' images appeared in only a few original tweets in our data. 
For example, image (a) appeared in 15 tweets. These tweet's total retweet count based on the Twitter metadata~(as returned from the API) counts add up to 24,399. The images were retweeted~(as recorded in our dataset) from users in the~\promoter~cluster 18,020 times. 
We note that images (a) and (b) were also the top two images retweeted by users in the \textit{suspended} users set, with 5,547 and 3,122 retweets in that set, respectively (recall that almost all suspended users belong to the \promoter~cluster). %

As expected, the pattern of retweeted images in the \detractor~cluster is quite different.
The three most retweeted images in the \detractor~cluster (not included for lack of space) have somewhat lower spread, appearing in tweets that were retweeted 10743, 6425, and 3411 times (based on metadata). The top image is a screenshot of the NY Times front page of Nov 11th, 2020 reporting that top election officials across the country have not identified any fraud. 

\begin{figure*}[t!]
\centering
\fontsize{9}{10}\selectfont
\begin{tabular}{lr|lr}
\toprule
\multicolumn{2}{c}{\promoter~cluster}&
\multicolumn{2}{c}{\detractor~cluster}\\
Domain&Retweets&Domain&Retweets\\
\midrule
pscp.tv&51,822             &washingtonpost.com&11,220 \\
youtube.com&44,031         &rawstory.com&9,267        \\
thegatewaypundit.com&35,967&cnn.com&4,139             \\
davidharrisjr.com&18,793   &independent.co.uk&3,882   \\
foxnews.com&17,332         &nytimes.com&3,746         \\
theepochtimes.com&15,297   &newsweek.com&3,496        \\
thedcpatriot.com&14,958    &news.yahoo.com&2,899      \\
thefederalist.com&13,288   &deadstate.org&2,409       \\
djhjmedia.com&11,816       &theguardian.com&2,232     \\
justthenews.com&11,149     &politicususa.com&2,032    \\
\bottomrule
\end{tabular}
\captionof{table}{
Top 10 domains being retweeted in the \promoter~and the \detractor~clusters respectively, as well as the number of retweets by users in these clusters.}
\label{tab:topdomains}
\end{figure*}

The analysis presented above can be easily extended with less-strict image similarity matching by calculating the Hamming distance between a pair of perceptual hash values. 
In this initial analysis, we used a strict sense of similarity, treating images as similar only when they share the exact same perceptual hash values.

\paragraph{URLs.}
We conduct preliminary analyses of the external links that have been included in the \datasetname~tweets.
\tabref{tab:topdomains} lists the top 10 domains that have been shared inside the \detractor~and \promoter~clusters respectively.
Most of the links shared by users in the \detractor~clusters are to mainstream news media, such as the Washington Post, CNN, 
and the New York Times.
The rest are other news-related websites.
The links shared by users in the \promoter~cluster mostly point to less authoritative news sources. 
Through manual inspection,
we find that 50\% of the top domains shared by the \promoter~cluster were evaluated as low-quality news sources by either MediaBiasFactCheck\footnote{mediabiasfactcheck.com} or by \citet{grinberg2019fake}.

The most shared domain in the \promoter~cluster is pscp.tv,
a live video streaming app owned by Twitter. 
YouTube stands out as the second most retweeted domain among the \promoter~users.
This trend is reflected in multiple news reports, warning of the significant role that YouTube plays in spreading false information related to voter fraud claims~\cite{superspreader}.
The majority of the top 10 most retweeted videos by the \promoter~users falsely claim evidence of widespread election fraud.
The users spreading these videos had significant overlap with the subsequent suspension action by Twitter.
For eight of the top 10 most retweeted videos by the \promoter~cluster, 29\%-42\% of the retweets of tweets sharing those videos were by accounts later suspended by Twitter. %

A scan of the top 10 YouTube channels retweeted in the \promoter~cluster shows that they were relatively large (millions of subscribers),
though there are also several smaller channels. 
For example,
the most retweeted channel, Precinct 13, has only 3.67K subscribers,
but has a video that appeared in 88 tweets and has been retweeted over 9K times.

Despite YouTube's announcement that it will take actions against content creators who falsely claim the existence of widespread voter fraud\footnote{see:twitter.com/YouTubeInsider/status/1347231471212371970},
as of Jan 11th,
the top 10 channels and videos listed in our tables are still available on YouTube.

\section{Related Work and Datasets}
We review prior work using Twitter data analysing politically related events,
with an emphasis on those that have released a public dataset.

In particular, prior research had heavily used and published Twitter data to study U.S. elections. 
Using tweets collected during the 2016 U.S. election,
researchers have analysed information operations run by social bots~\cite{rizoiu2018debatenight},
characterized the dissemination of misinformation~\cite{vosoughi2018spread} and the exposure of American voters to misinformation~\cite{grinberg2019fake}.
Work in \citet{hua2020characterizing,hua2020towards} characterized adversarial interaction against political candidates during the 2018 U.S. general election and shared 1.7M~tweets interacting with political candidates. 

Focusing on the U.S. 2020 election,
research studied false claims regarding mail-in ballots~\cite{benkler2020mail} before the election as the COVID-19 pandemic made it hard to vote in person.
Closest to our work is the \textit{\#Election2020} dataset~\cite{chen2020election2020},
which streamed a broad set of Twitter data for both political candidates' tweets and related keywords.
As discussed above,
although some of the voter fraud related keywords were included in their data collection process,
our \datasetname~dataset contains more than 2.3~times as much of the related data in \textit{\#Election2020}, 
for the `voter fraud' keyword,
presumably because of our more focused stream. 
Our stream also included a broader set of fraud-claim related keywords. 

In order to help understand the dissemination of misinformation across platforms,
\citet{brena2019news,hui2018hoaxy} used news articles as queries and released the tweets pointing to these articles.
In 2018, Twitter published a list of accounts that the platform suspects to be related to Russia's government controlled Internet Research Agency~\cite{twitterira}.
This release enabled a number of studies that deepened our understanding of foreign information manipulation in the U.S.~\cite{arif2018acting,im2020still,badawy2018analyzing}.

Most of the previous works that released Twitter datasets only included the tweet IDs, in accordance with Twitter's Terms of Service.
We keep to that practice, and augment the data without sharing tweet content, as detailed above,
making our multi-modal dataset more accessible and useful to the research community.

\section{Discussion and Conclusions}

The unsubstantiated voter fraud claims spread on Twitter and elsewhere around the U.S. 2020 presidential elections are likely to form one of the most consequential misinformation campaigns in modern history. 
It is critical to allow a diverse set of researchers to provide a deeper understanding of this effort, which will continue to have national and global impact for years to come. 
To enable that contribution, it is important to provide a public and accessible archive of this campaign on various social media platforms, including Twitter as we do in \datasetname. 

The \datasetname~dataset has the potential to benefit the research community, 
and to further inform the public regarding both Twitter user activities around the voter fraud claims, as well as Twitter's own response. 
Yet, the data has some limitations. 
We could not possibly capture the full extent of the voter fraud claims on Twitter, as our dataset was constructed by using matching keywords. 
Further, as analyzed above, we do not have full coverage even for the keywords we tracked, though we estimate that we have a majority of the tweets with those keywords. 
Nevertheless, the breadth of the data enables various types of investigation using both the tweet data, as well as the aggregated data of URLs, videos and images used in the campaign. We propose three major categories of such investigation. 

First, researchers can use the dataset to study the spread, reach, and dynamics of the voter fraud campaign on Twitter. 
Researchers can describe and analyze the participants, including the activities of political candidates using information from orthogonal datasets of candidate accounts~\footnote{For example, \url{https://github.com/vegetable68/Midterm-2020-candidates}}, or the interaction between public figures and other accounts spreading claims and promoting certain narratives.
Further, the data can help expose how different public figures spread different claims, for example the claims regarding the Dominion voting machines, what kind of engagement such narratives received.
The data can also be used to understand the role of bots and other coordinated activities and campaigns in spreading this information. 
In general, the dataset can provide for analysis of the distribution of attention to these claims and how it spreads -- via images, tweets, URLs -- including comparison among different pre-computed communities and clusters. 

Second, we include auxiliary data -- URLs including YouTube links, and image hashes -- that can help researchers examine \textit{other} sources of information and their role in spreading the voter fraud claims. For example, using the image hash values that were encoded using publicly available algorithms, researchers can easily map images not just within the Twitter data, but also from the larger web and media ecosystem. 
Researchers may combine our dataset with datasets that are collected from other social media platforms to examine how visual misinformation spread across platforms~(e.g., \cite{zannettou2018origins,moreira2018image}).

A third potential area of investigation is Twitter's response to this campaign of spreading voter fraud claims. 
Twitter's civic policy and in particular its approach to misinformation around the election results and claims of voter fraud had been rated as ``non-comprehensive'' and was generally not applied consistently and transparently \cite{CenterforanInformedPublic2021}.  
The \datasetname~dataset can help better understand Twitter's response. A specific question is the characterization of the suspended users whom as we shown above are primarily part of a specific community, related to QAnon. 
Researchers can use the data to both understand Twitter's non-public response and its effectiveness, or even simulate the effectiveness of hypothetical earlier bans of the same population or users. 
As noted above, while Twitter's terms of service do not allow us to publicly sharing full data for the suspended users---the \datasetname~tweets for these users are no longer available on Twitter by their ID---we will make these tweets available privately to published academic researchers, as we believe these tweets are of immense and justified public interest. 

The publicly released \datasetname~data was collected and made available according to Twitter's Terms of Service for academic researchers, following established guidelines for ethical Twitter data use~\cite{rivers2014ethical}. 
By limiting to the Tweet IDs as the main data element,
the dataset does not expose information about users whose data had been removed from the service. 
The only content in our data that is directly tied to a Tweet ID is the hash of the images for tweets that included them. Even though that hash, theoretically, can be tied to an image from another source, in absence of the original tweet the image will not be associated with any user account. 
We believe that this minor disclosure risk is justified given the potential benefits of this data.

\section*{Acknowledgements}

We thank the anonymous reviewers for their helpful feedback.
This research was funded by NSF research grant CNS-1704527.

\clearpage
\appendix
\section*{Appendix}
\label{ap:hashtags}
\begin{table}[ht]
\fontsize{9}{10}\selectfont
\centering
    \begin{tabular}{lrr}
    \toprule
    \multicolumn{3}{c}{\textbf{Community 0}}\\
    Handle  & Active Status & Retweets \\
    \midrule
    kylegriffin1 & active & 76,302\\
    mmpadellan & active & 74,393\\
    donwinslow & active & 69,796\\
    BernieSanders & active & 60,961\\
    AriBerman & active & 58,222\\
    \midrule
    \multicolumn{3}{c}{\textbf{a) Community 1}}\\
    \midrule
    realDonaldTrump & suspended & 1,560,373\\
    LLinWood & suspended & 1,057,805\\
    SidneyPowell1 & suspended & 633,273\\
    GenFlynn & suspended & 334,197\\
    CodeMonkeyZ & suspended & 274,210\\
    \midrule
    \multicolumn{3}{c}{\textbf{b) Community 2}}\\
    \midrule
    DonnaWR8 & suspended & 38,388\\
    zeusFanHouse & suspended & 36,347\\
    LeahR77 & suspended & 33,352\\
    TheRISEofROD & suspended & 32,992\\
    Bubblebathgirl & active & 27,787\\
    
    \midrule
    \multicolumn{3}{c}{\textbf{c) Commmunity 3}}\\
    ganaha\_masako & active & 12,480\\
    KadotaRyusho & active & 6,890\\
    yamatogokorous & active & 5,716\\
    mei98862477 & active & 5,347\\
    kohyu1952 & active & 5,244\\
    \midrule
    \multicolumn{3}{c}{\textbf{d) Community 4}}\\
    FernandoAmandi & active & 4,217\\
    POTUS\_Trump\_ESP & active & 2,981\\
    TDN\_NOTICIAS & active & 2,459\\
    1VAFI & active & 1,802\\
    Gamusina77 & active & 1,638\\
    \bottomrule
    
    \end{tabular}

\caption{\label{fig:top-users-by-community} Top 5 Users in each community sorted by retweets from other users. }
\end{table}

\begin{table}[!t]
\fontsize{9}{10}\selectfont
\centering
\hyphenpenalty=10000
\begin{tabular}{p{62 mm}}
\textbf{Seed list}\\
\hline
\#abolishdemocratparty \#ballotharvasting \#ballotvoterfraud \#cheatingdemocrats
\#democratvoterfraud
\#gopvoterfraud
\#ilhanballotharvesting
\#ilhanomarballotharvesting
\#ilhanomarvoterfraud
\#mailinvoterfraud
\#stopvoterfraud
\#voterfraud
\#voterfraudbymail
\#voterfraudisreal\\
\\
\textbf{Filtered out}\\
\hline
\#abolishdemocratparty \\
\\
\textbf{Generated from the seed list}\\
\hline
\#ballotharvesting
\#voterid
\#ilhanomarforprison
\#stopgopvoterfraud
\#ilhanomar
\#nancypelosiabusingpower
\#nancypelosimustresign
\#junkmailballots
\#traresforcongress
\#immigrationfraud
\#votebymailfraud
\#ballotfraud
\#exposed
\#votersuppression
\#ilhanresign
\#voteinperson
\#votebymail
\#video
\#lockherup
\#nomailinvoting
\#ilhanomarelectionfraud
\#taxfraud
\#ballotharvesting
\#massivemailinballots
\#arrestilhanomar
\#obamagate
\#ilhanomarlockherup
\#buyingvotes
\#2020election
\#campaignfraud
\#homewrecker
\#voteinperson
\#minneapolis
\#absenteeballots
\#trump2020
\#arrestilhanomar
\#absenteeballot
\#darktolight
\#wwg1wga
\#terrorist
\#daveygravyspirualsavage
\#trump
\#fraud
\#liar
\#pizzagate
\#republicans
\#qproof
\#theawakening
\#voteatthepolls
\#marriedherbrother
\#glasshouses
\#sheepnomore
\#voteyouout
\#cheater
\#georgesoros
\#georgia
\#vote
\#walkaway
\#thegreatawakening
\#qanon
\#evil
\#savethechildren\\
\\
\textbf{Keywords list 10/24} \\
\hline
\#ballotfraud
\#ballotharvesting
\#ballotvoterfraud
\#cheatingdemocrats
\#democratvoterfraud
\#ilhanomarballotharvesting
\#ilhanomarvoterfraud
\#mailinvoterfraud
\#nomailinvoting
\#stopgopvoterfraud
\#stopvoterfraud
\#votebymailfraud
\#voterfraud
\#voterfraudisreal\\
\\
\textbf{Added on 10/31} \\
\hline
\#discardedballots
\#electionfraud
\#electioninterference
\#electiontampering
\#gopvoterfraud
\#hackedvotingmachines
`destroyed ballots'
`discarded ballots'
`election fraud'
`election interference'
`election tampering'
`hacked voting machine'
`pre-filled ballot'
`stolen ballots'
`ballot fraud'
`ballot harvesting'
`cheating democrats'
`democrats cheat'
`harvest ballot'
`vote by mail fraud'
`voter fraud'\\
\\
\textbf{Added on 11/03} \\
\hline
\#stopthesteal \\

\end{tabular}
\caption{\label{tab:hashtags} Hashtags and keywords related to election fraud.}
\end{table}

\clearpage
{\bibliography{citation}}

\end{document}